\documentclass[amsmath,amssymb,aps,floatfix,prb,twocolumn]{revtex4-1}
\usepackage{graphicx}
\begin{document}
\title{Exotic collective dynamics in molten Carbon}

\author{Taras Bryk$^{1,2}$, Giancarlo Ruocco$^{3,4}$, 
Jean-Fran{\c c}ois Wax$^{5}$, No\"el Jakse$^{6}$}

\affiliation{ $^1$ Institute for Condensed Matter Physics,National
Academy of Sciences of Ukraine,\\UA-79011 Lviv, Ukraine}
\affiliation{$^2$Institute of Applied Mathematics and Fundamental
Sciences,\\Lviv National Polytechnic University, UA-79013 Lviv,
Ukraine } 
\affiliation{$^3$
Center for Life Nano- \& Neuro-Science, Istituto Italiano di
  Tecnologia, 295 Viale Regina Elena, I-00161, Roma, Italy}
\affiliation{$^4$ Dipartimento di Fisica, Universita' di
Roma "La Sapienza", I-00185, Roma, Italy} 
\affiliation{$^5$ Laboratoire de Chimie et de Physique A2MC, Universit\'e de Lorraine,
        Metz, F-57000 Metz, France}
\affiliation{$^6$ Universit\'e Grenoble Alpes, CNRS, Grenoble INP, SIMaP, F-38000  Grenoble, France}

\date{\today}
\begin{abstract}
Collective longitudinal and transverse propagating modes in molten Carbon at $T=5500$~K  and pressure range $10$-$40$ GPa are reported from {\it ab initio} based as well as machine learned molecular dynamics. 
A striking exotic feature in collective dynamics is the two-peak shape of the current spectral functions of the single-component liquid, which makes evidence of a second branch of longitudinal propagating modes in the wave number range $k>1$\AA$^{-1}$. 
It is shown that time correlation functions reflecting the out-of-phase motion of particles and their cages of nearest neighbors results in the same frequencies of the exotic low-frequency branch of vibrations.
A theoretical framework of generalized collective modes is applied to recover the time dependence of density-density, imaginary part of susceptibility and longitudinal current-current correlations.
\end{abstract}

\maketitle

Molten Carbon, existing only at extremely high temperatures, still poses scientific challenges  due to the complex bonding versatility and resulting rich and competing structural motifs, as well as its importance for instance for the study of carbon-rich planetary interiors \cite{Hul20}.  
It therefore triggers active experimental\cite{Kra13, Knu08, Pri20, Kra25} and simulation\cite{Gal89,Gal90, Dha90,Bet20} studies, revealing features of its microscopic structure and single-particle dynamics.  
Very recent X-ray diffraction and X-ray free-electron laser experiments reported the structure factor 
of liquid C in the temperature range $6500$-$7300$~K \cite{Kra25}, revealing a four-coordinated liquid displaying transient bonding consistent with quantum-based simulations.

Collective dynamics in one-component liquids is quite simple and well understood on macroscopic spatial scales, when the liquid is treated as continuum. 
Then, collective dynamics is represented as a collection of propagating and relaxing collective modes, life time of which scales with wave number $k$ as $\sim k^{-2}$, that is a consequence of local conservation laws \cite{Han}. 
Outside the hydrodynamic regime, the non-hydrodynamic processes caused by fluctuations of non-conserved quantities contribute to the observed time-dependent correlations, and theoretical description should account for the existence of non-hydrodynamic modes \cite{deS88,Bry11}.

Atomistic computer simulations enable numerical estimation of time-dependent correlations in liquids in a very wide range of wave numbers $k$, far outside the hydrodynamic regime.
A general feature of collective dynamics of one-component liquids outside the hydrodynamic region is a presence of the only extended longitudinal acoustic propagating mode with dispersion law $\omega(k)$ having a linear hydrodynamic form $\omega(k\to 0)=c_sk$ in the long-wavelength region. 
The dispersion law of propagating collective modes is usually obtained from computer simulations \textit{via} peak position $\omega_{max}(k)$ of longitudinal (L) current spectral functions $C^L(k,\omega)$, which can be obtained by numerical Fourier-transformation of current-current time correlation functions derived by molecular dynamics (MD) simulations. 
So far no observations from MD simulations were reported about liquids having two well-defined maxima of $C^L(k,\omega)$, although experimental data from IXS experiments suggested the existence of two contributions from collective propagating modes in liquid Bi and Sb \cite{Inu15,Inu21}. 
Another on-going discussion is on the possibility of transverse (T) shear waves to contribute to the longitudinal dynamics \textit{via} possible L-T coupling due to anisotropy of the first coordination shell \cite{Hos09,Hos13}. 
Furthermore, {\it ab initio} MD (AIMD) simulations of transverse dynamics in several different liquids under pressure revealed a two-peak structure of the transverse current spectral functions $C^T(k,\omega)$ \cite{Bry15,Bry20}. 
So far, the origin of the high-frequency transverse branch from these simulations remains elusive. 

The two-peak structure of the longitudinal current spectral functions exists in binary liquids as it was shown in \cite{Bry00}. In long-wavelength region the $C^L(k,\omega)$ has a single peak corresponding to the hydrodynamic acoustic excitations, while outside the hydrodynamic regime another peak emerges due to the non-hydrodynamic
optic mode \cite{Bry02}.
\begin{figure*} [t]
\includegraphics[width=0.32\textwidth]{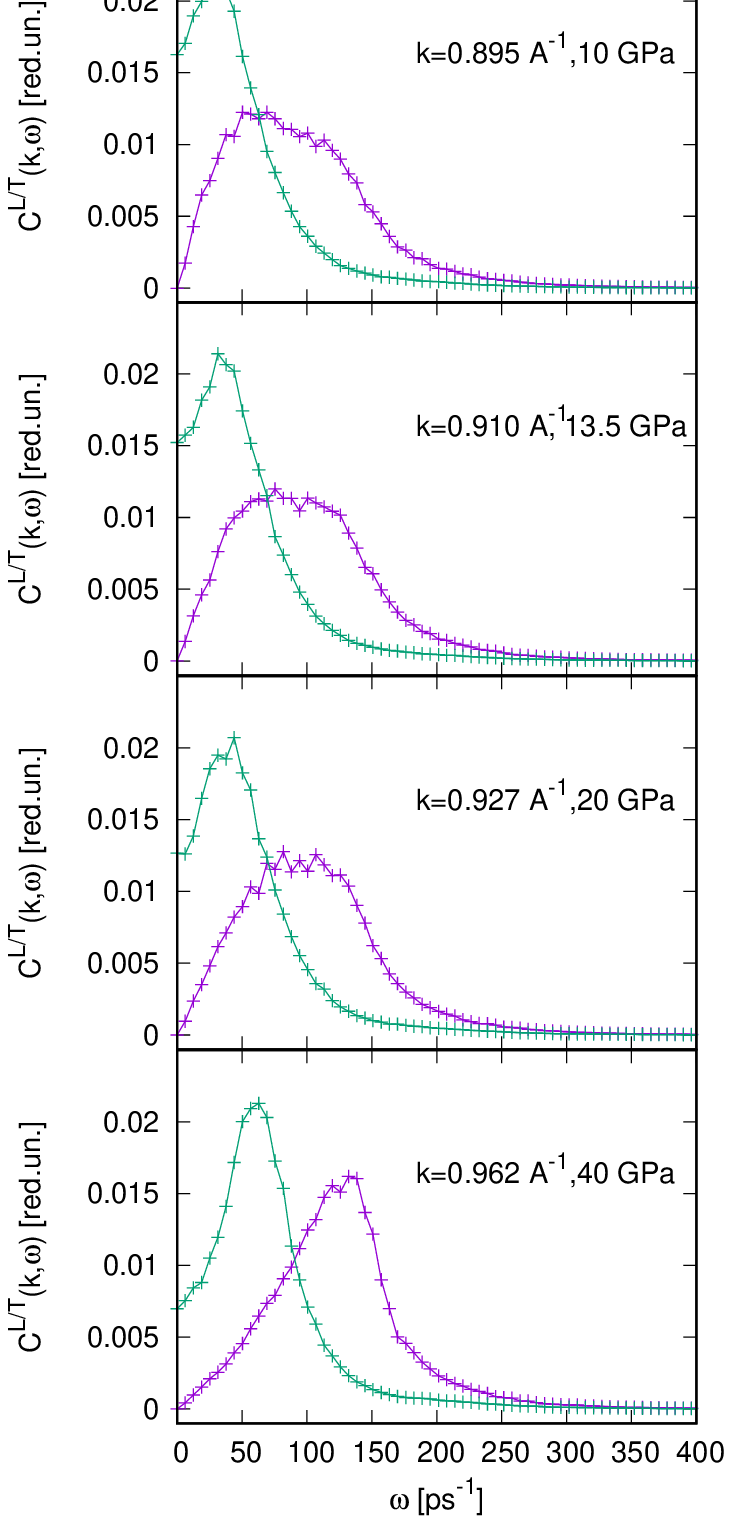}%
\includegraphics[width=0.32\textwidth]{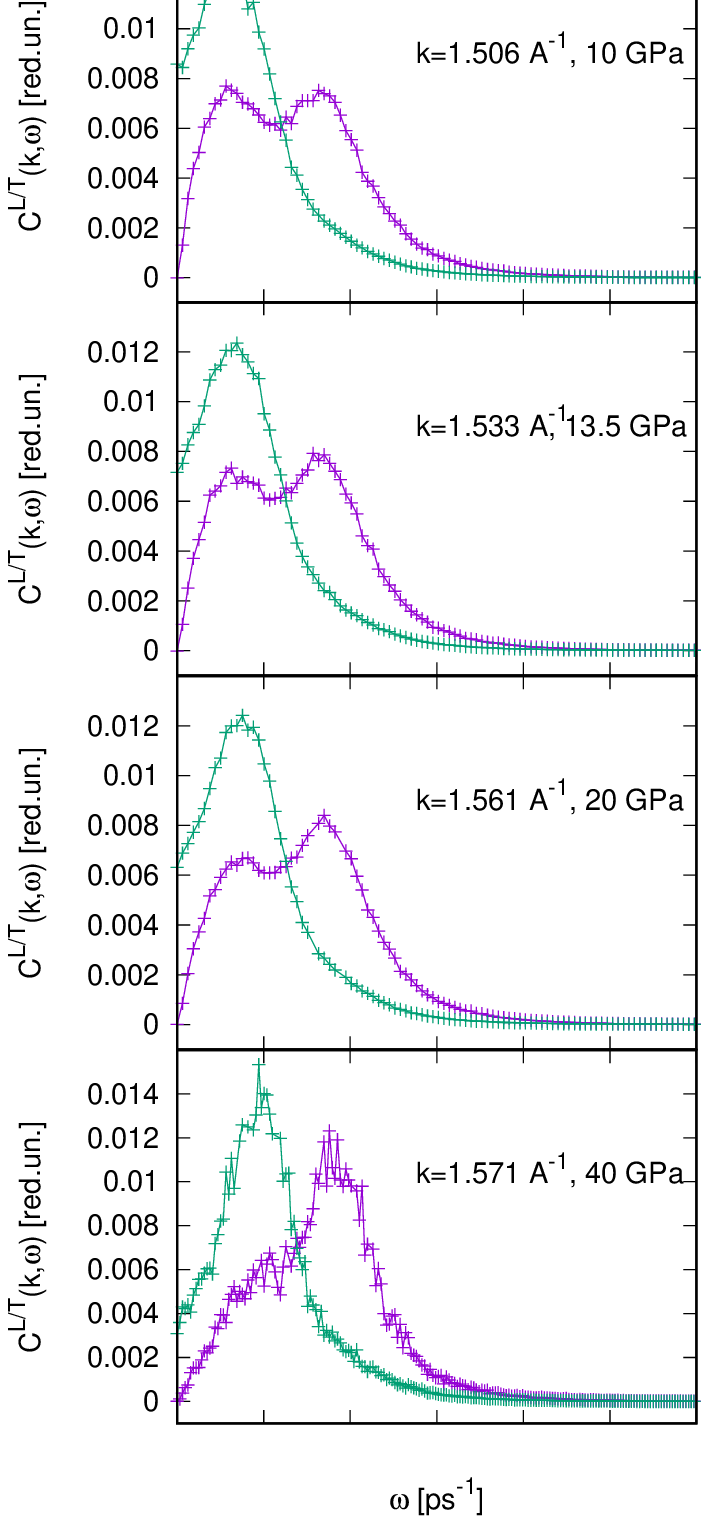}%
\includegraphics[width=0.32\textwidth]{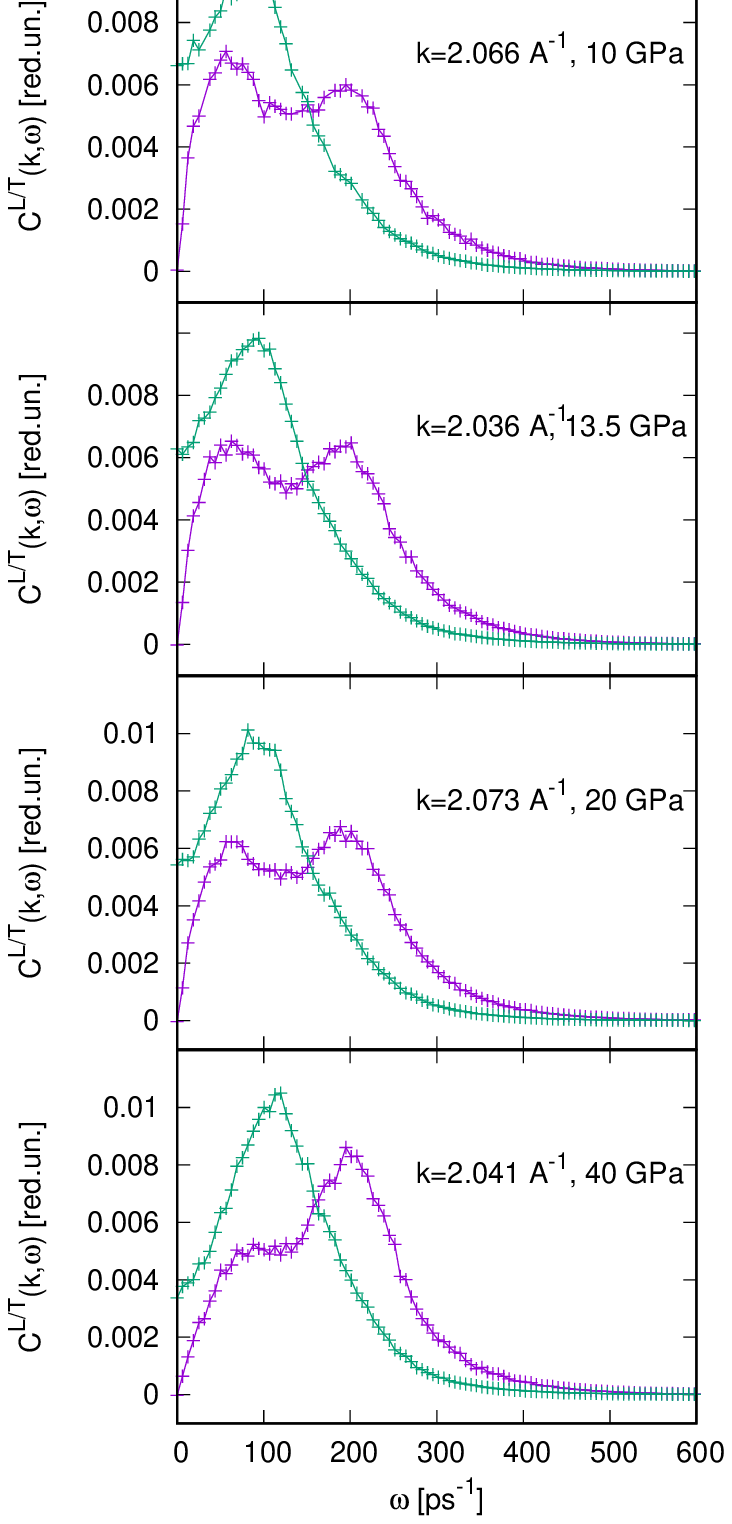}%
\caption{Longitudinal and transverse current spectral 
functions $C^{L/T}(k,\omega)$ at three wave numbers and four pressures 
for molten Carbon at $T=5500$~K, obtained
from {\it ab initio} simulations.} \label{cltkw4p}
\end{figure*}

\begin{figure}
\includegraphics[width=0.45\textwidth]{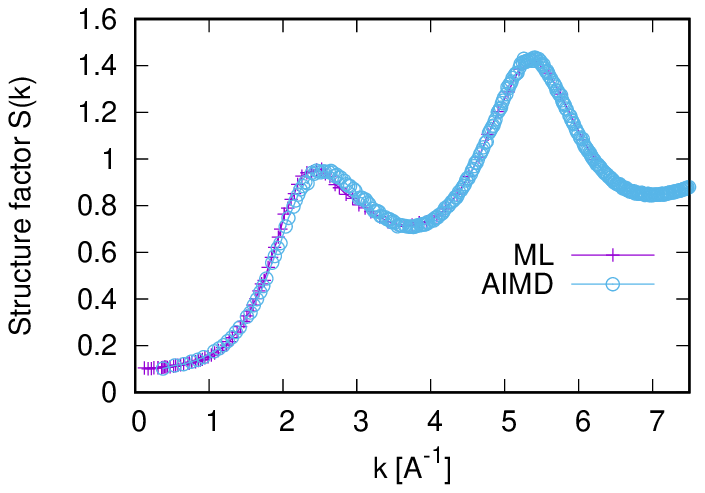}%
\caption{Structure factor for liquid C at T=$5500$~K at $P=20$~GPa from ML simulations and AIMD.
}
\label{skC}
\end{figure}

\begin{figure}
\includegraphics[width=0.45\textwidth]{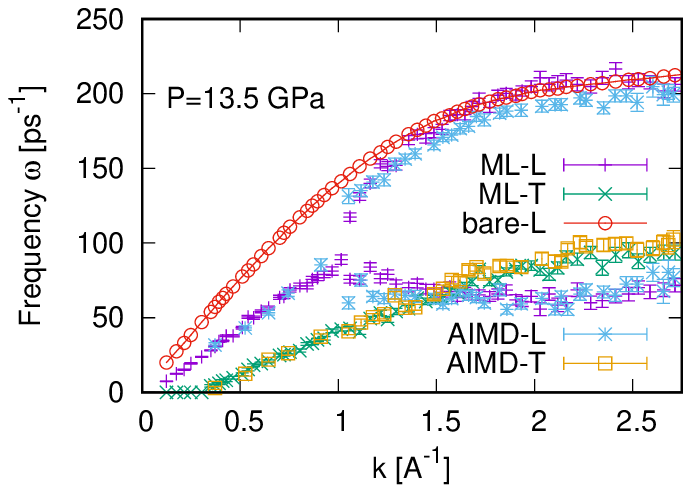}%

\includegraphics[width=0.45\textwidth]{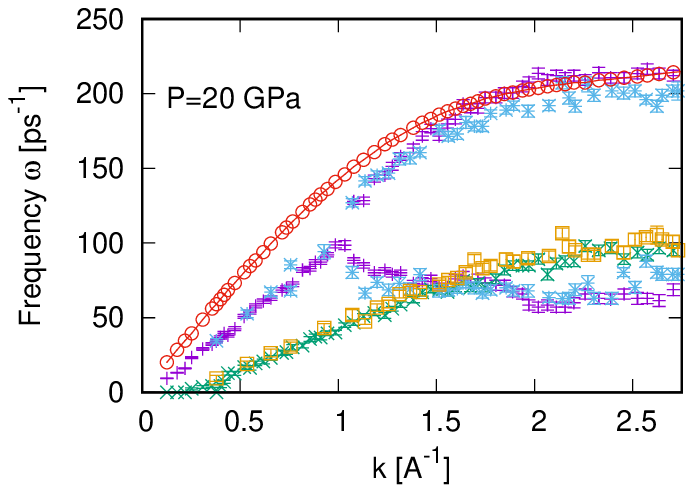}%
\caption{Peak positions of L and T current spectral functions for
liquid C at $5500$~K and pressures $13.5$ and $20$~GPa from ML simulations and AIMD
}
\label{dispC}
\end{figure}

The aim of this Letter is to report an exotic two-peak structure of the longitudinal current spectral functions $C^L(k,\omega)$, never observed before for one-component liquids to the best of our knowledge.
For this purpose, AIMD simulations were performed within the density function theory (DFT) for molten Carbon with $600$ atoms using the Vienna \textit{Ab initio} Simulation Package (VASP)\cite{Kresse1996}. 
Temperature $5500$~K and four different pressures resulted in the L and T time correlation functions in the Fourier space, as shown in Fig.\ref{cltkw4p}. 
The details of simulations are provided in Supporting information file \cite{SupI}.
In order to make sure that the results of {\it ab initio} simulations were not affected by the size effects especially in the small wave-vector regime, a general purpose Machine Learning Interatomic Potentials (MLIP) was trained on ab initio phase-space trajectories using the High-Dimensional Neural Network (HDNN) scheme of Behler and Parrinello \cite{Behler2007,Behler2021}. 
With this potential MD simulations with $16200$ atoms were performed with the \textsc{Lammps} code \cite{Lammps} and the \texttt{hdnnp} package \cite{Singraber2019} (called hereafter ML simulations). 
It was shown recently that the HDNN was able to reproduce with \textit{ab initio} accuracy the structure and dynamics monatomic liquids \cite{Jakse2023,Demmel2025} including complex elements such as boron \cite{Sandberg2024}. 
As a matter of fact, our MLIP perfectly recovered the static structure of liquid carbon as can be seen on Fig.\ref{skC}) as well as  frequencies of collective excitations on Fig.\ref{dispC}) from {\it ab initio} simulations.

\begin{figure*}
\includegraphics[width=0.45\textwidth]{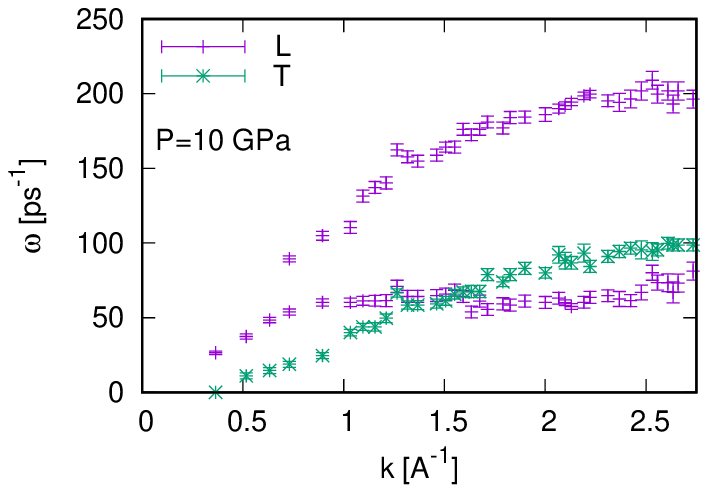}%
\includegraphics[width=0.45\textwidth]{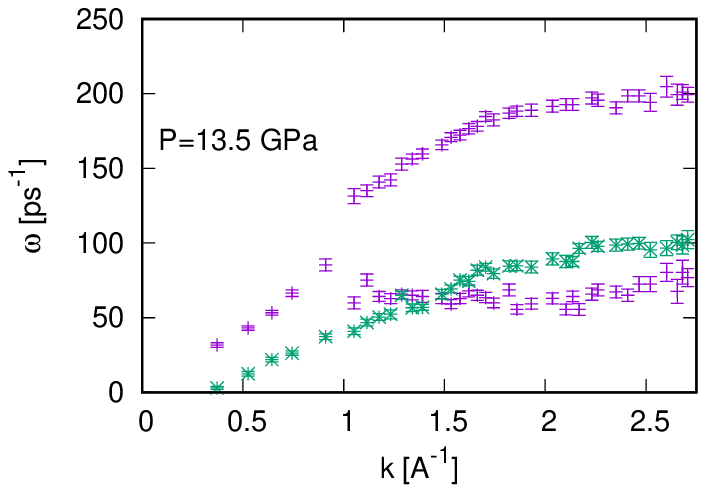}%

\includegraphics[width=0.45\textwidth]{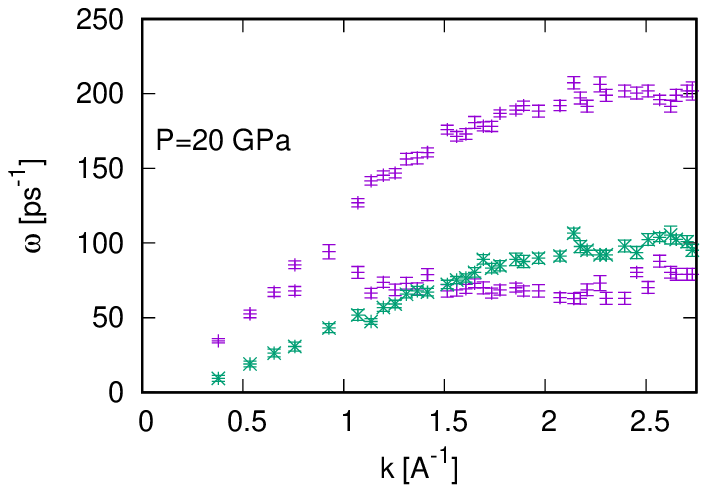}%
\includegraphics[width=0.45\textwidth]{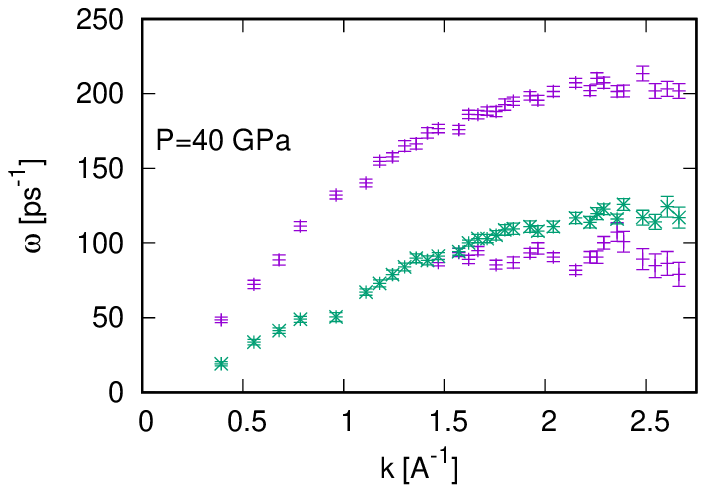}%
\caption{Observed maxima positions of the longitudinal and transverse current spectral 
functions $C^{L/T}(k,\omega)$ for molten C at four pressures and temperature 5500~K, obtained
from {\it ab initio} simulations.} \label{disp}
\end{figure*}

Fourier spectra of which for several $k$-values are shown in Fig.\ref{cltkw4p}. 
One can see that for small wave numbers both $C^{L/T}(k,\omega)$ spectral function have just a single-peak shape, and the peak location corresponds to the frequencies of acoustic (L) and shear (T) waves in molten Carbon. 
For higher wave numbers we clearly observed an unusual two-peak shape (Fig.\ref{cltkw4p}) of the $C^L(k,\omega)$ and a single-peak one in the transverse case.
The dispersion of L and T excitations at four studied pressures, obtained from the observed maxima of L and T current spectral functions,
are shown in Fig.\ref{disp}.


One can see in Fig.\ref{dispC} that for $k>1$\AA$^{-1}$ a split of the numerically obtained dispersion of acoustic modes into two branches. 
Another conclusion can be made that the low-frequency L branch does not recover the dispersion of the shear waves (T branch). 
For $k>1$\AA$^{-1}$ the high-frequency branch definitely corresponds to the high-frequency acoustic modes, which follows from its comparison with the dispersion of "bare" (non-damped) acoustic modes
shown in Fig.\ref{dispC} by open circles. However, the origin of the low-frequency L branch for $k>1$\AA$^{-1}$ is unclear from such a comparison.

\begin{figure}
\includegraphics[width=0.48\textwidth]{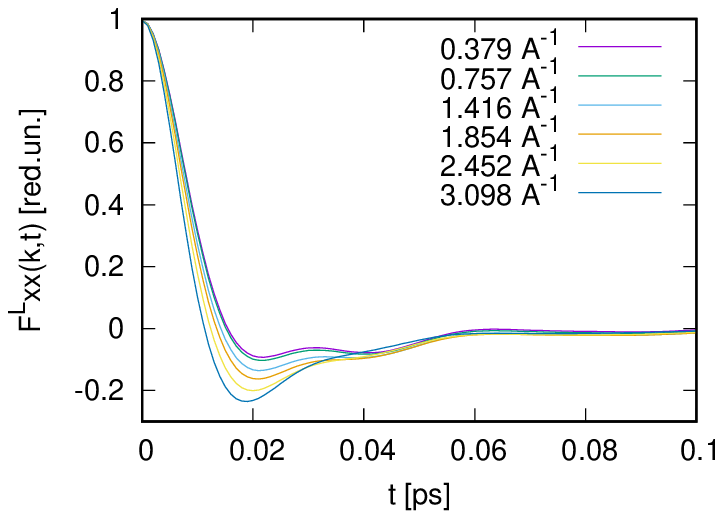}%

\includegraphics[width=0.48\textwidth]{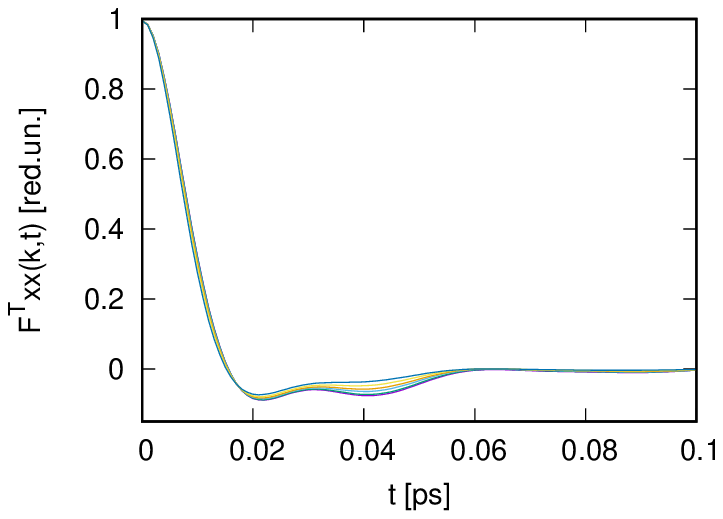}%
\caption{Longitudinal and transverse autocorrelation functions $F^{L/T}_{xx}(k,t)$ 
of out-of-phase 
motion of particles and their cages for molten C at 5500~K and P=20.6 GPa.
} \label{fxx}
\end{figure}

\begin{figure}
\includegraphics[width=0.45\textwidth]{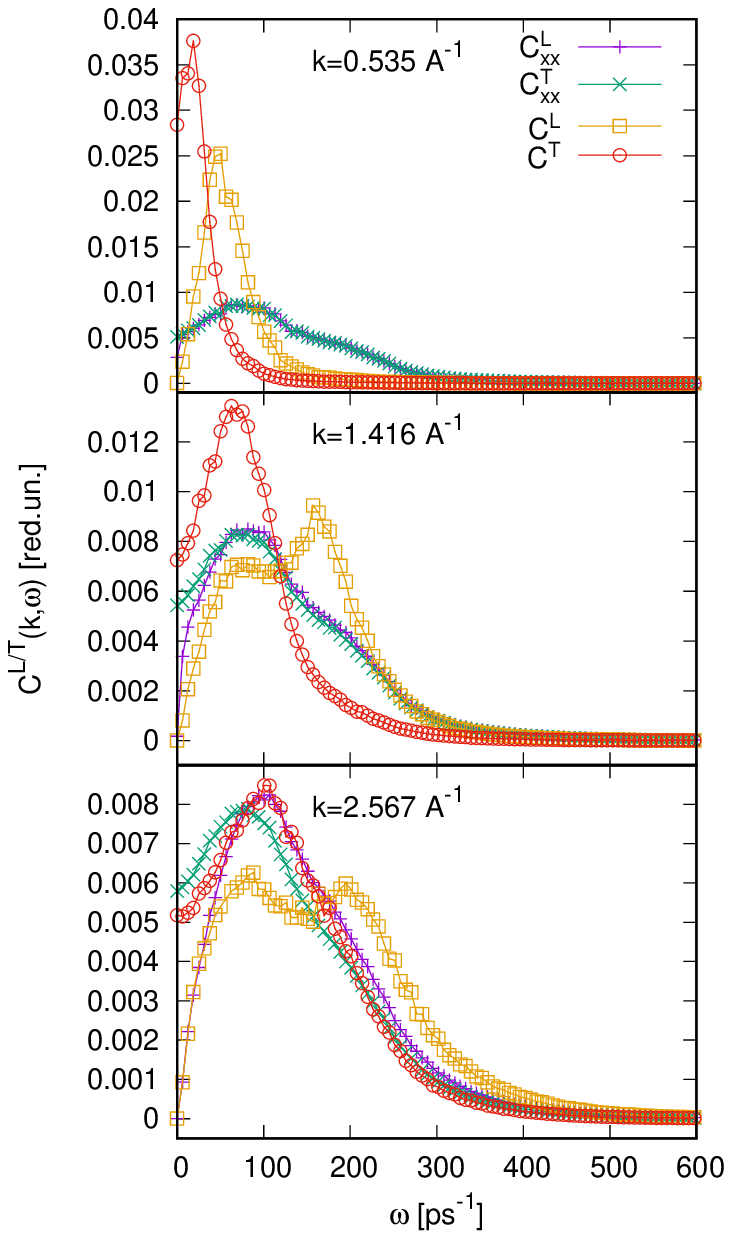}%
\caption{Spectral functions of total L (open boxes) and T (open circles) currents
and mutual L (plus symbols) and T (cross symbols) currents of particles and their cages 
for molten C at $5500$~K and $P=20$ GPa.
} \label{clt}
\end{figure}

Now the task is to understand the origin of the high- and low-frequency longitudinal branches at $k>1$\AA$^{-1}$. The high-frequency branch is simply the extension of the acoustic  dispersion, that follows from their comparison with the dispersion of acoustic "bare" 
modes (see Supplement Informations file \cite{SupI}). 
In order to estimate the origin of the non-acoustic low-frequency L branch of collective excitations, we took into account that the non-acoustic branch in binary liquids comes from the fluctuations of mutual (opposite  motion of two kinds of particles in a binary system) current. 
Another point is that in liquids collisions of particles in fact cause the out-of-phase motion of neighboring particles. Therefore,
we suggest to sample from the simulation data a dynamic variable ${\bf J}_j(k,t)$, which corresponds to the mutual current of a $j$-th particle and its all neighbors, as follows. 
\begin{equation}
{\bf J}_j(k,t)=\frac{1}{N}
 [(N-1){\bf v}_j e^{-i{\bf kr}_j}- \sum_{j'\neq j}^N {\bf v}_{j'}e^{-i{\bf kr}_{j'}}]~.
\label{Jj}
\end{equation}
Here, the summation is performed over all the particles other than the selected $j$-th one. Note that the form (\ref{Jj}) guarantees that during the sampled window of MD configurations all out-of-phase motions of particles and their instantaneous neighbors will be taken into account even when different particles form the first coordination shell due to the diffusive processes. 
Outside the first coordination shell the correlation in the out-of-phase rapidly drops to zero because of the absence of long-range stable cages due to diffusivity. 
It is important that the dynamic variable ${\bf J}_j(k,t)$ is orthogonal to the total current density fluctuations ${\bf J}(k,t)$, i.e.
$$
\langle  J^{L/T}_j(k,t_0)J^{L/T}(-k,t_0)\rangle_{t_0,j}\equiv 0~,
$$ 
meaning that the dynamic variable (\ref{Jj}) describes dynamic processes which the hydrodynamic one ${\bf J}(k,t)$ cannot. 
The longitudinal (L) and transverse (T) components of this currrent and corresponding L and T current-current time correlation functions were calculated in a standard way
$$
F^{L/T}_{xx}=\langle  J^{L/T}_j(k,t)J^{L/T}_j(-k,t=0)\rangle_{t_0,j}~,
$$
where the double average was performed over different $j$-particles and time origins.
The time dependence of these L and T sime correlation functions at several $k$-values is shown in Fig.\ref{fxx}.
Remarkably enough, the spectral functions $C^{L/T}_{xx}(k,\omega)$ shown in Fig.\ref{clt} both have only a single maximum, which correlates very well with the low-frequency peak of the L and T spectral functions of total current. 

\begin{figure*}[t]
\includegraphics[width=0.45\textwidth]{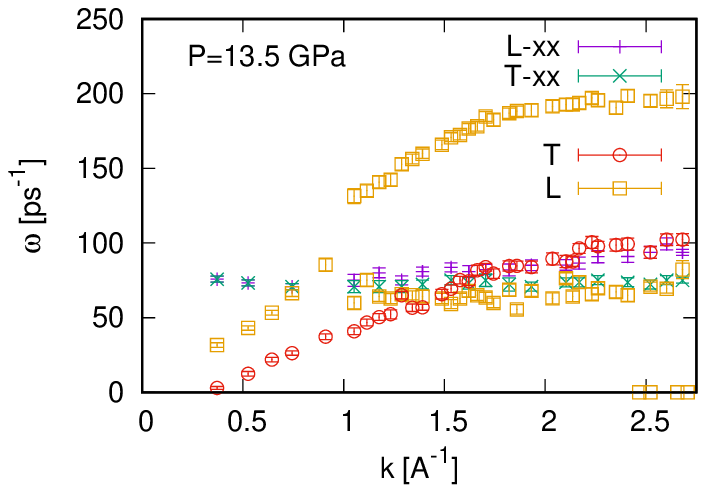}%
\includegraphics[width=0.45\textwidth]{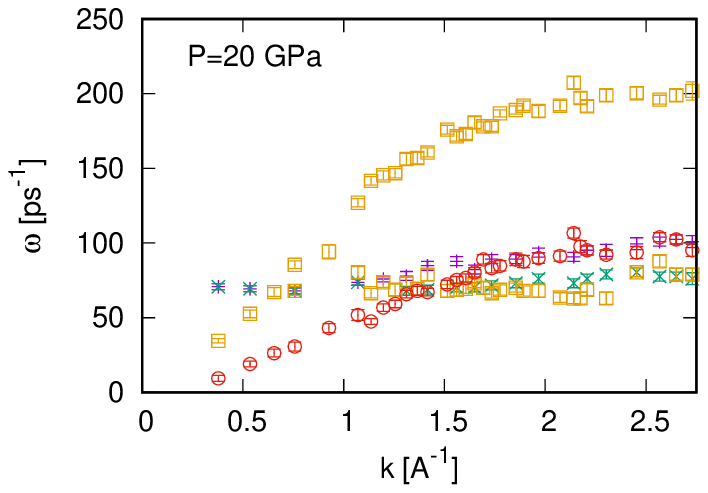}%
\caption{Dispersion of longitudinal (open boxes) and transverse (open circles) 
collective excitations 
from maxima positions of AIMD-derived current spectral functions $C^{L/T}(k,\omega)$
and maxima of spectral functions of mutual currents $C^{L/T}_{xx}(k,\omega)$ (plus 
and cross symbols) for molten C at P=13.5 GPa and 20 GPa.
} \label{dispxx}
\end{figure*}

In Fig.\ref{dispxx} we show the positions of the maxima of the spectral functions of mutual currents $C^{L/T}_{xx}(k,\omega)$ (plus and cross symbols) together with the dispersion of the collective modes obtained from the total current spectral functions for molten C at two
pressures $P=13.5$ GPa and $20$ GPa.
It is well seen that the maxima of $C^{L/T}_{xx}(k,\omega)$ correspond right to the dispersion curves of the exotic nonacoustic low-frequency branch of collective excitations (see Fig.\ref{dispxx}).
Now it follows that similar as this is in the binary liquids the out-of-motion of carbon atom and its neighbors cause the non-hydrodynamic optic-like branch, which is not revealed at low wave numbers in the shape of total current spectral functions (Fig.\ref{cltkw4p}), but contribute to the collective dynamics at microscopic length-scale $l\sim 4-7$~\AA. Obviously, this is related to the existence of medium-range order in the molten C (see Fig.\ref{skC}).

In summary, AIMD and ML simulations of molten Carbon at four pressures in the range $10$-$40$~GPa reveal an exotic two-peak shape of the current spectral functions, which results in two high- and low-frequency branches of propagating modes for $k>1$\AA$^{-1}$. 
The high-frequency branch is a standard dispersion of the high-frequency sound, while the low-frequency one corresponds to some non-hydrodynamic propagating modes. 
In order to estimate the origin of the low-frequency branch of L-excitations, several types of time correlation were probed between fluctuations of non-conserved quantities. 
The spectral functions $C^{L/T}_{xx}(k,\omega)$ of the out-of-phase motion of particles and their cages of their nearest neighbors were found to have single low-frequency maxima, restoring the low-frequency peak of $C^L(k,\omega)$.  

{\it Acknowledgment}.
TB was supported by the National Research Foundation of Ukraine under Project 2023.05/0019. NJ acknowledges the CINES, TGCC and IDRIS under Project No. INP2227/72914/gen5054, as well as CIMENT/GRICAD for computational resources. This work has been partially supported by MIAI@Grenoble Alpes (ANR-19-P3IA-0003). Discussions within the French collaborative network in artificial intelligence in materials science GDR CNRS 2123 (IAMAT) are also acknowledged.

\newpage

\section*{Supporting Information}

\section*{Ab initio Molecular Dynamics Simulations}
{\it Ab initio} molecular dynamics (AIMD) simulations of molten carbon were performed by the Vienna
Ab initio Simulation Package (VASP) \cite{Kre93,Kre96,Kre96b} with a system of 600 particles 
under periodic boundary conditions in the three dierection of space. 
The electron-ion interaction was represented by PAW-potentials \cite{Blo94,Kre99} and exchange-correlation 
effects were treated by generalized gradient approximation in Perdew-Burke-Ernzerhof (PBE) version \cite{Per96}. 
The plane wave expansion cutoff was $400$~eV. 
Only $\Gamma$ point of the Brillouin zone was taking in the sampling of electron density.

The dynamic simulations were performed by solving Newton's equations numerically using Verlet's algorithm in the velocity form with a time step of $1$~fs. 
The phase-spece trajectories were built in the canonical ensemble, NVT  (constant number of atoms, $N$,) constant volume $V$, and constant temperature, $T$) was used to control the mean temperature of the simulations at $5500$~K by means of a Nose-Hoover thermostat. 
Molten Carbon was simulated at four pressures: $10$, $13.5$, $20$ and $40$~GPa by adjusting initially the size of simulation box
followed by an equlibration over 10 ps for each pressure. 
Runs were then continued of 80 ps for the production of the physical properties at equilibrium.   

\section*{Machine Learning Interatomic Potential and classical simulations}

\subsection*{Ab initio Dataset}

An essential aspect of the construction of a machine learning interatomic potential (MLIP) is to build a dataset for training and testing that contains configurations for the various phases in a sufficient temperature and pressure range so that the potential probes the configuration space dedicated to the study under consideration. 
Following the methodology of our previous work \cite{Jakse2023}, the dataset gathers configurations obtained by sampling AIMD trajectories with a regular sampling each $10$ time steps that results in $21282$ configurations in total \textit{i.e.} $12.7\times10^6$ local environments. This tight sampling guaranties that fluctuations of energy are taken into account, that is the tail of the energy distribution are also sufficiently sampled.
The goal is primarily to describe liquid carbon, but basic crystalline phases were also included, leading to a general purpose MLIP.
In addition to the $4$ thermodynamic states used in this work, additional shorter AIMD simulations of typically $10$ ps to $20$ ps are performed as described above.
For the liquid state, temperatures of $4500$~K, $5000$~K, $5500$~K, $6000$~K, $6500$~K were considered. 
For the crystalline phases, the high pressure diamond structure at $0$~K and $1000$~K and graphite structure at $3000$~K were simulated.   In each case, various pressures were considered for the purpose of probing the repulsive part of the potential.

\subsection{High-Dimensional Neural Network approach}

MLIP such as the one built with HDNN framework are based on a nearsightedness principle for which the total energy $E_{\text{total}}$ of $N$ atoms is the sum of atomic energies, namely 
\begin{equation}
	E_{\text{total}} = \sum_{i=1}^{N} E_i,
\end{equation}
where $E_{i}$ is the local energy of atom $i$. In the framework of neural networks it is exploited as follows
\begin{equation}
	E_i = f_\theta(D_i),
\end{equation}
with $f_{\theta}$, a function depending on the set of parameters $\theta$ of the neural network architecture. 
The descriptor $D_{i}$ of the central atom $i$ is a feature vector representing its local atomic environment. 
In the present work, the Behler-Parrinello atom centered symmetry functions (ACSF) \cite{Behler2011} were used. 
They are comprised of radial or angular terms describing respectively the distance and angle distributions between the atom $i$ and its surrounding neighbors, and possess the transnational and rotational invariance by definition. 
The following radial function
\begin{equation}
	G_i^2 = \sum_{j \neq i} e^{-\eta (R_{ij} - R_s)^2} f_c(R_{ij})
\end{equation}
and angular functions 
\begin{equation}
	\begin{split}
		G_i^5 = 2^{1-\zeta} \sum_{j \neq i} \sum_{k \neq i, j} \left( 1 + \lambda \cos \theta_{ijk} \right)^\zeta  e^{-\eta \left( R_{ij}^2 + R_{ik}^2 + R_{jk}^2 \right)} \\
		f_c(R_{ij})  f_c(R_{ik})  f_c(R_{jk})
	\end{split}
\end{equation}
were considered in the present work. 
$R_{ij}$ represents distance between central atom $i$ and atom $j$, and $\theta_{ijk}$ represents the angle between the atom $i$ and two neighboring atoms $j$ and $k$. 
$f_c(R_{ij})$, $f_c(R_{ik})$ and $f_c(R_{jk})$ are cutoff functions beyond which they tend smoothly to zero and interactions between atoms are neglected, thus defining the spherical local environment. 
The parameters $\eta$, $\zeta$, $\lambda$, and $R_s$ are given a set values to cover the radial and angular information of the local atomic environment. 

A number of $12$ radial functions and $48$ angular functions were considered with a cutoff radius $R_C=5.5$ \AA{}, resulting to a $60$-dimensional feature space. For the radial functions, values of $\eta = 1.2356E-02$, $3.3387\times10^{-2}$, $9.1298\times10^{-2}$, $2.4817\times10^{-1}$, $6.7461\times10^{-1}$, $1.8338$, and $3.6675$ \AA{}$^{-1}$. Values of $R_S = 0$, $7.5714\times10^{-1}$, $1.5143$, $2.2714$, $3.0286$, $3.7857$, and $4.5429$ \AA{}. For the angular functions, the values of $\lambda =1$ and $-1$. The values of $\zeta = 1$, $2$, $4$, $8$, $16$, and $64$. Values of $\eta = 0.01$, $0.12$, and $0.5$ \AA{}$^{-1}$. Values of $R_S = 0$ and $4.85$ \AA{}. 

The dataset was then split into two subsets: $90$ percent for training and $10$ percent for testing. 
The test set was subsequently used to evaluate the model performance through the Root-Mean-Square Error (RMSE).
The training was performed solely on energies \cite{Jakse2023} using a set of multi-layer perceptron neural networks with $2$ hidden layer of $25$ neuron each.
Optimization of the weights and bias are performed using Adam stochastic gradient method and a standard Kalman filter was used. A RMSE of 9.17 meV/atom is obtained. 

\begin{figure*}
	\includegraphics[width=0.90\textwidth]{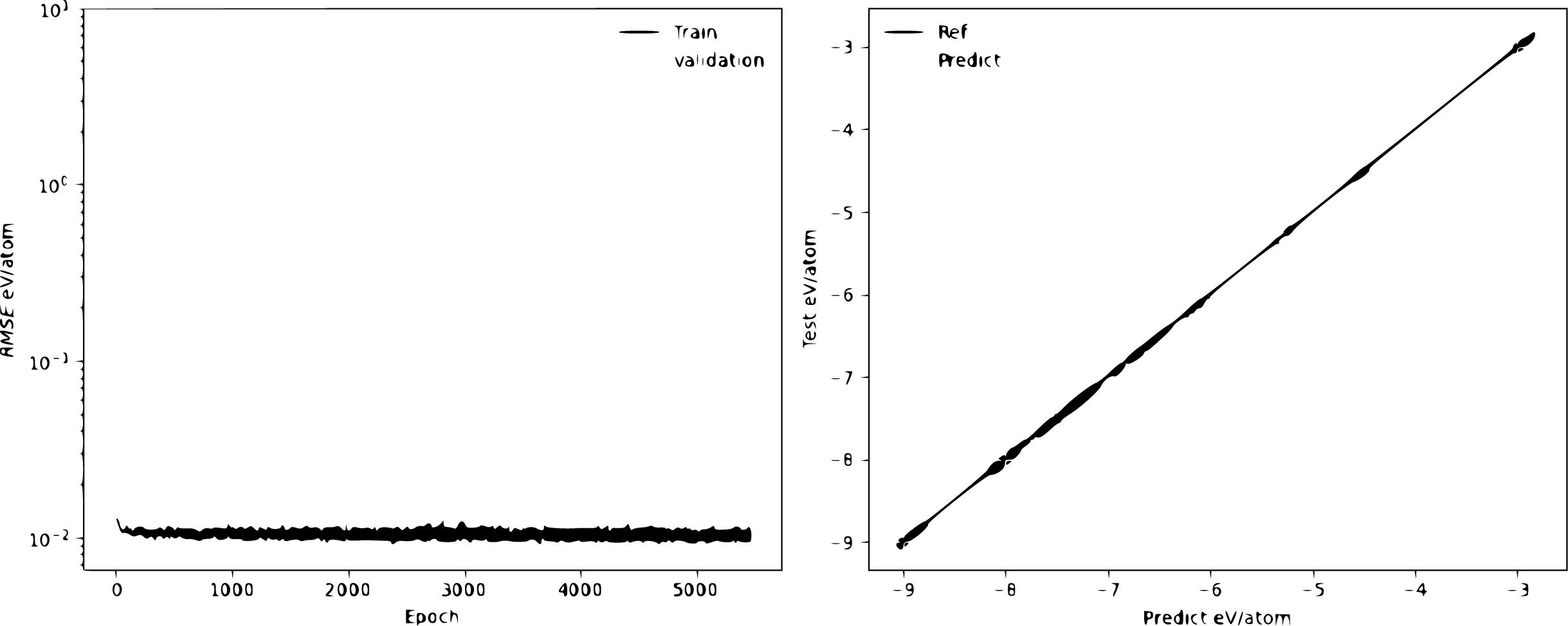}%
	\caption{Learning curves (left panel) and Test-Predict diagram (right panel), where the blue line represents a perfect prediction.} \label{fig-learning}
\end{figure*}

The learning curve with per atom RMSE values of the training and the validation on the test set are drawn in Fig. \ref{fig-learning}, where the final potential is taken with weights and bias corresponding to the minimum of this curve.  

\section*{Static and single-particle dynamic properties of molten Carbon from {\it ab initio} molecular dynamics simulations}

\begin{figure*}
	\includegraphics[width=0.48\textwidth]{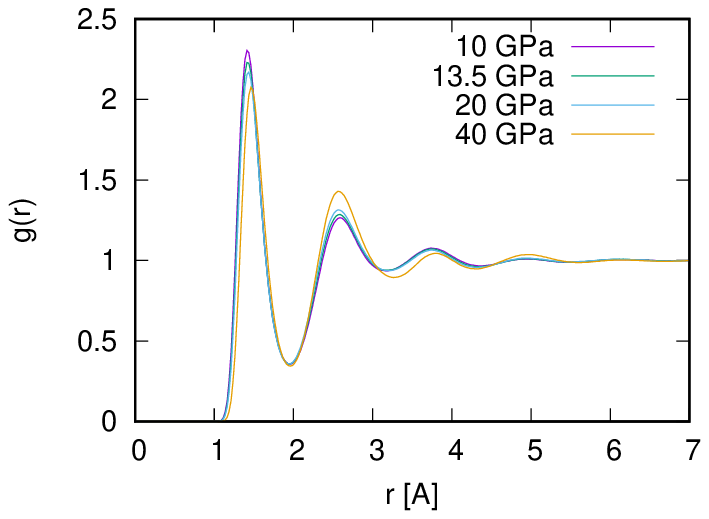}%
	\includegraphics[width=0.48\textwidth]{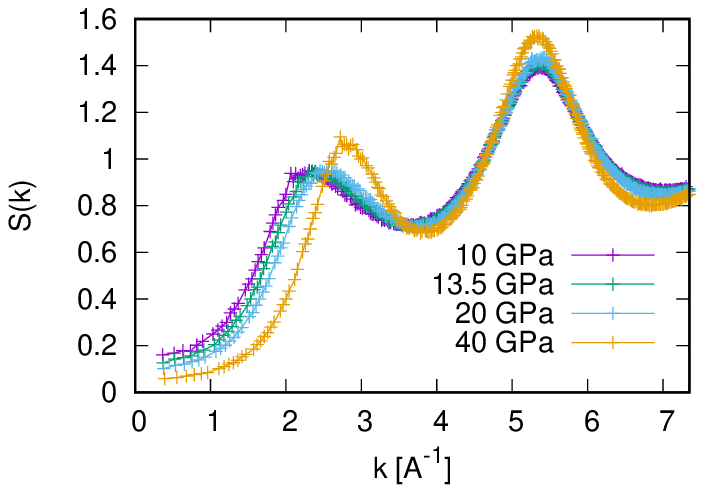}%
	\caption{Pair distribution functions $g(r)$ and static structure factors $S(k)$ 
		for molten Carbon at
		four pressures and temperature 5500~K, obtained from {\it ab initio}
		simulations.} \label{str}
\end{figure*}

Fig.\ref{str} represents the pair distribution functions (PDF) $g(r)$ and the corresponding static structure factors at the four studied pressures in AIMD. The pair distribution functions are typical as are in one-component liquids, and the effect of pressure is in reduction of the amplitude of the main maximum of PDF and increase of the amplitude of the second maximum. The static structure factors $S(k)$ which were calculated as instantaneous density-density correlators 
$$
S(k)=\langle n_{-k}n_k\rangle~,
$$
where $n_k(t)$ are the spatial-Fourier components of particle density, 
show a typical for molten C prepeak, which changes its localition with 
pressure from $k\sim 2$\AA$^{-1}$ at P=10~GPa to $k\sim 2.8$\AA$^{-1}$ at P=40~GPa, while the location of main peak of $S(k)$ remains practically unchanged with pressure. The existing prepeak of $S(k)$ is an evidence of partial chemical ordering of Carbon atoms.  

\begin{figure*}
	\includegraphics[width=0.48\textwidth]{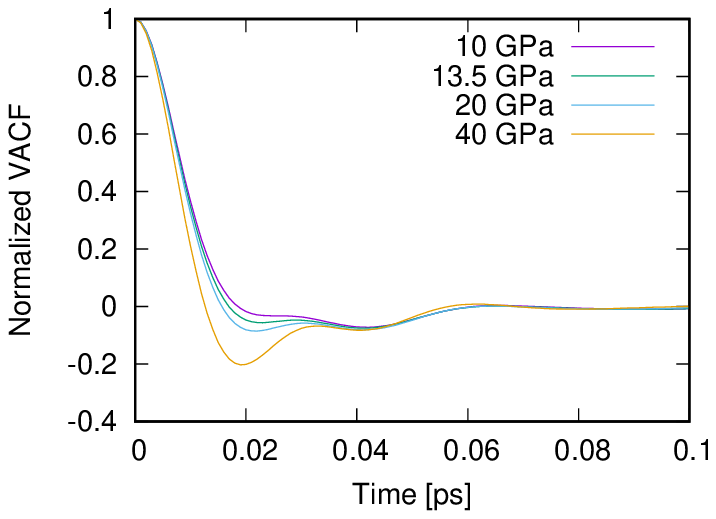}%
	\includegraphics[width=0.48\textwidth]{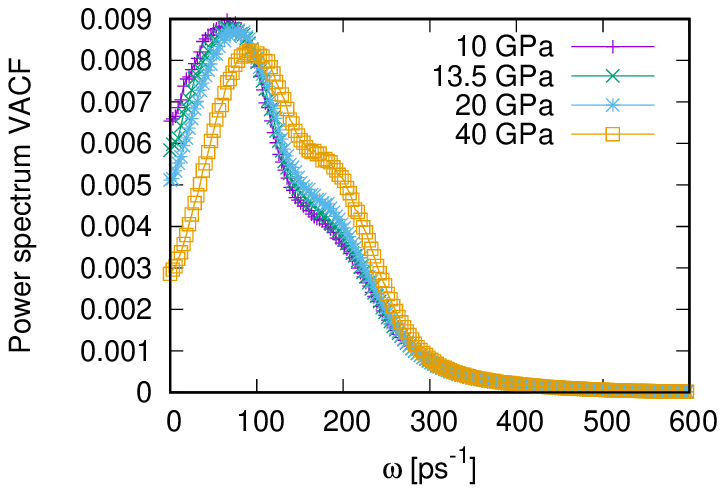}%
	\caption{Normalized velocity autocorrelation functions and their Fourier spectra 
		for molten Carbon at
		four pressures and temperature 5500~K, obtained from {\it ab initio}
		simulations.} \label{vacf}
\end{figure*}

The simplest single-particle velocity autocorrelations in molten Carbon show a gradual change with pressure (see Fig.\ref{vacf}). For all the studied pressure the Fourier spectrum of velocity autocorrelation functions show a well-defined low-frequency maximum, changing with pressure from $\omega_{max}\sim 50$~ps${-1}$ to $\sim 100$~ps${-1}$, and a shoulder in the 
region of frequencies $\sim 170-200$~ps${-1}$. 

\section{Theoretical analysis of time correlation functions of molten Carbon obtained from {\it ab initio} molecular dynamics simulations}

Analysis of the AIMD-derived time correlation functions in terms of dynamic eigenmodes was performed within the approach of Generalized Collective Modes (GCM) \cite{Mry95,Bry10} using the methodology suggested in \cite{Bry13}. The GCM eigenmodes were obtained from the $5\times 5$
generalized hydrodynamic matrix ${\bf T}(k)$ generated for each $k$-point sampled in AIMD using the set of
$5$ dynamic variables of the thermo-viscoelastic (TVE) dynamic model \cite{Bry10,Bry13}
\begin{equation} \label{a5}
	{\bf A}^{(TVE)}(k,t) = \left\{n(k,t), J^L(k,t), \varepsilon(k,t),
	\dot{J}^L(k,t), \dot{\varepsilon}(k,t)\right\}.
\end{equation}
Here the spatial-Fourier components of particle density are
\begin{equation} \label{dynhyd1}
	n(k,t)=\frac{1}{\sqrt{N}}\sum_{j=1}^{N}e^{-i{\bf kr}_j}~,
\end{equation}
and the spatial-Fourier components of the mass-current density were 
\begin{equation}
	{\bf J}(k,t)=\frac{m}{\sqrt{N}}\sum_{i=1}^{N}{\bf v}_i(t)e^{-i{\bf kr}_i(t)}~,
\end{equation}
while their first time derivative similarly was
\begin{equation}
	{\dot {\bf J}}(k,t)=\frac{1}{\sqrt{N}}\sum_{i=1}^{N}
	[{\bf F}_i(t)+im({\bf kv}_i){\bf v}_i(t)]e^{-i{\bf kr}_i(t)}~,
\end{equation}
with $m$ being the atomic mass of Carbon, ${\bf r}_i(t)$, ${\bf v}_i(t)$ and ${\bf F}_i(t)$ were the particle trajectory, particle velocity and force acting on the i-th particle, respectively. The energy density and first time derivative of energy density in the TVE set of dynamic variables (\ref{a5}) were treated as recommended in Ref.\cite{Bry13}.

The eigenvalues of the generalized hydrodynamic matrix ${\bf T}(k)$ were either purely real (non-propagating relaxing eigenmodes) or pairs of complex-conjugated eigenvalues
$$z_{\alpha}(k)=\sigma_{\alpha}(k)\pm i\omega_{\alpha}(k)~,
$$
where $\sigma_{alpha}(k)$ is the $k$-dependent damping of the $\alpha$-th propagating mode and $\omega_{alpha}(k)$ - its dispersion.

\begin{figure*}
	\includegraphics[width=0.90\textwidth]{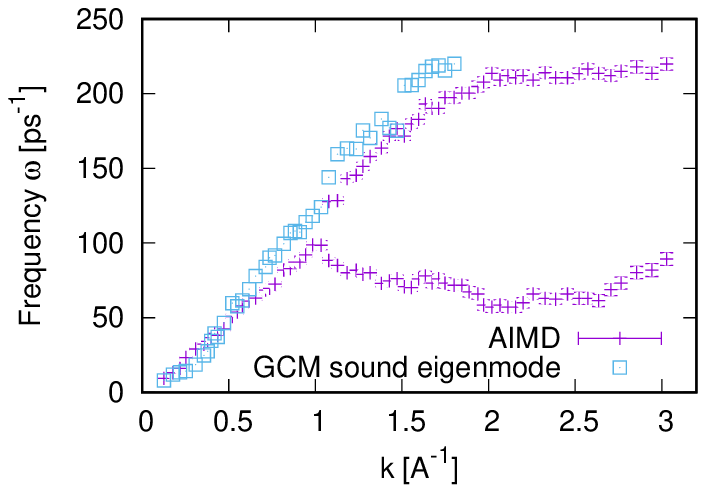}%
	\caption{GCM sound eigenvalues vs peak positions of L spectral functions for
		liquid C at 5500~K and pressure 20~GPa from ML simulations
	}
	\label{GCMz}
\end{figure*}

In Fig.\ref{GCMz} we show the imaginary part of the sound eigenvalues in a wide range of wave numbers that makes evidence that the high-frequency branch of propagating modes in molten C corresponds to propagating acoustic modes. The low-frequency branch in the region $k>1$\AA$^{-1}$ in Fig.\ref{GCMz} is definitely not of acoustic origin.

The ability of any dynamic model within the GCM approach for a correct description of collective dynamics in the simulated system can be verified by comparison of the AIMD-derived time correlation functions and their GCM expressions:
\begin{equation}\label{replica}
	F_{ij}(k,t)=\sum_{\alpha=1}^{N_{TVE}}G_{ij}(k)e^{-z_{\alpha}(k)t}~,~~i,j={\bf A}^{(TVE)}
\end{equation}
where the time correlation function between any two variables from the TVE set (\ref{a5})  are represented via the partial sum of five eigenmodes $z_{\alpha}(k)$ with weight coefficients $G_{ij}(k)$ expressed via the eigenvectors of corresponding eigenvalue \cite{Mry95}. 

\begin{figure*}
	\includegraphics[width=0.60\textwidth]{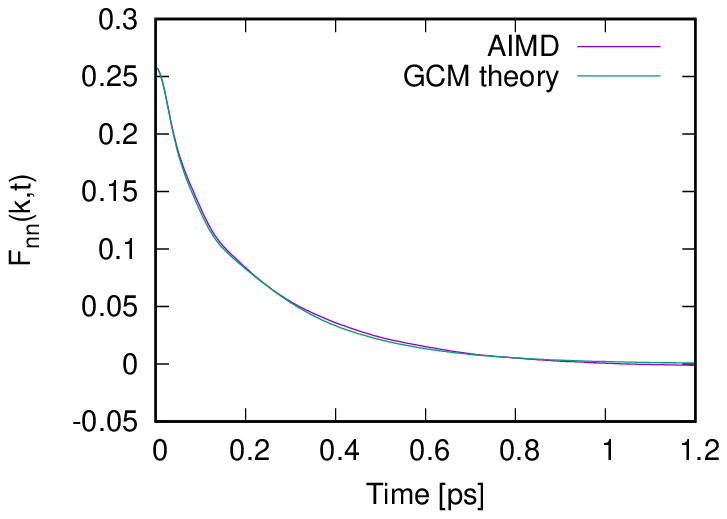}
	\includegraphics[width=0.60\textwidth]{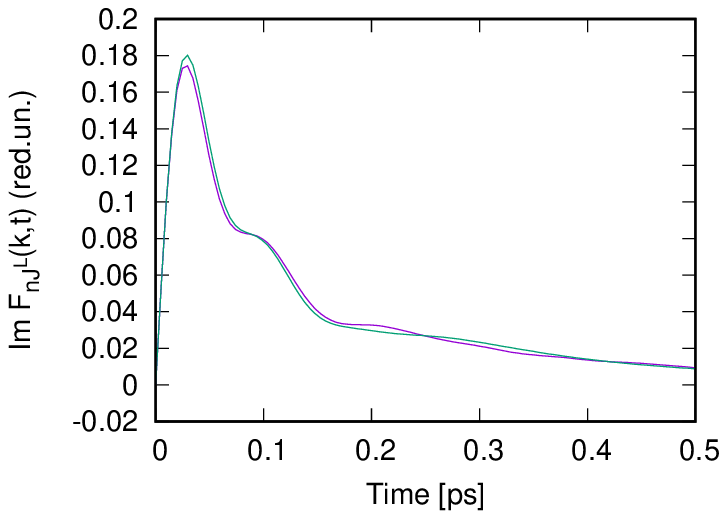}
	\includegraphics[width=0.60\textwidth]{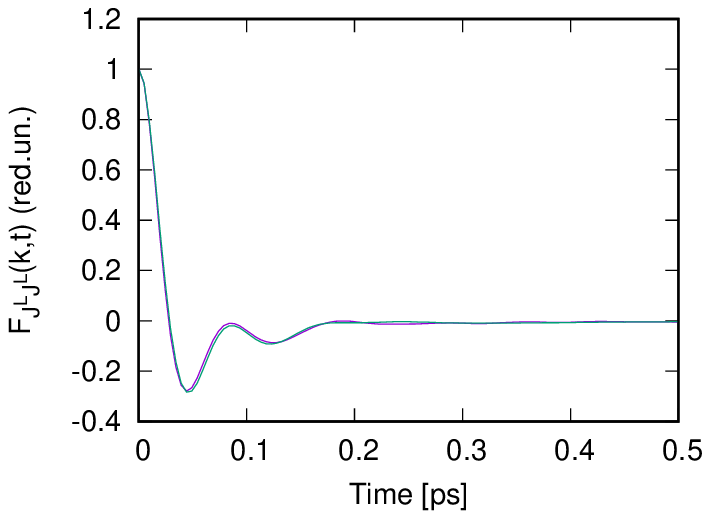}
	\caption{Quality of GCM theory in recovering density-density, imaginary part of density response and current-current 
		time correlations at k~1.4 A$^{-1}$ for liquid C at 5500~K and pressure 20~GPa 
	}
	\label{nntcf}
\end{figure*}

In Fig.\ref{nntcf} we show the good quality of the 5-variable TVE dynamic model in recovering the AIMD-derived time correlation functions of longitudinal collective dynamics in molten Carbon. Usually the GCM expression (\ref{replica}) was used to recover the density-density and current-current time correlation functions. In this AIMD study we additionally tried to recover another time correlation function: imaginary part of $F_{nJ}(k,t)$,
which is in fact the density response function.

\begin{figure*}
	\includegraphics[width=0.90\textwidth]{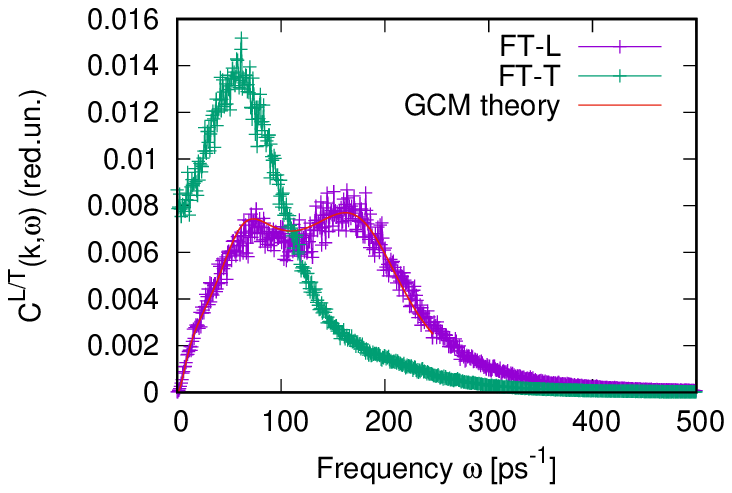}
	\caption{GCM theoretical current spectral function at k~1.4 A$^{-1}$ vs 
		L and T spectral functions from ML simulations
		for 
		liquid C at 5500K and pressure 20 GPa 
	}
	\label{CLKW}
\end{figure*}

The L- and T-current spectral functions are shown in Fig.\ref{CLKW}
at the same wave number as the time correlation functions in Fig.\ref{nntcf} for molten Carbon at pressure 20~GPa. The solid red curve
is the GCM spectral function of the longitudinal current-current time correlation function in Eq.(\ref{replica}), and it perfectly recovers the two-peak shape of the $C^L(k,\omega)$ spectral function derived from AIMD data. Each peak corresponds to a different longitudinal collective mode at $k=1.4$\AA$^{-1}$, frequencies of which form two branches of collective excitations in Fig.\ref{GCMz}.


\begin{thebibliography}{12}
%
\bibitem{Hul20} C.J. Hull, S.L. Raj, R.J. Saykally, Chem. Phys. Lett. {\bf 749} 137341 (2020)
%
\bibitem{Wi2012} H. F. Wilson and B. Militzer,  The Astrophysical Journal, 745(1), 54 (2012).
%
\bibitem{Kra13} D. Kraus et al , Phys. Rev. Lett. {\bf 111} 255501 (2013)
%
\bibitem{Knu08} M. D. Knudson, M. P. Desjarlais, D. H. Dolan. Science, {\bf 322} 1822 (2008)
%
\bibitem{Pri20} E. Principi, S. Krylow, M. E. Garcia, A. Simoncig, L. Foglia, R. Mincigrucci, 
G. Kurdi, A. Gessini, F. Bencivenga, A. Giglia, S. Nannarone, C. Masciovecchio  , Phys. Rev. Lett. {\bf 125} 155703 (2020)
%
\bibitem{Kra25} D. Kraus et al, Nature (2025) https://doi.org/10.1038/s41586-025-09035-6
%
\bibitem{Gal89} G. Galli, R.M. Martin, R. Car, M. Parrinello, Phys. Rev. Lett. {\bf 63} 988 (1989)
%
\bibitem{Gal90} G. Galli, R.M. Martin, R. Car, M. Parrinello, Phys. Rev. B {\bf 42} 7470 (1990)
%
\bibitem{Dha90} M. W. C. Dharma-wardana, F. Perro, Phys. Rev. Lett. {\bf 65} 76 (1990)
%
\bibitem{Bet20}  M. Bethkenhagen, B. B. L. Witte, M. Sch\"orner, Gerd R\"opke, Tilo D\"oppner,  D. Kraus, S. H. Glenzer, P. A. Sterne, R. Redmer, Phys. Rev. Research {\bf 2} 023260 (2020)
%
\bibitem{Han}   J.-P.Hansen and I.R.McDonald, {\it Theory of Simple
                Liquids} (London: Academic) (1986).
\bibitem{deS88} I.M.~de~Schepper, E.G.D.Cohen, C.Bruin, J.C.~van~Rijs,
                W.Montfrooij, and L.A.~de~Graaf, Phys. Rev. A {\bf 38},
                271 (1988).                
%
\bibitem{Bry11} T.Bryk, Eur. Phys. J. Spec. Top., {\bf 196}, 65 (2011)
%
\bibitem{Inu15} M. Inui, Y. Kajihara, S. Munejiri, S. Hosokawa, A. Chiba, K. Ohara, S. Tsutsui,
                A.Q.R. Baron.
                Phys.Rev. B {\bf 92}, 054206 (2015).
%
\bibitem{Inu21} M. Inui, Y. Kajihara, S. Hosokawa, A. Chiba, Y. Nakajima Y., et al.
                J. Phys.: Condens. Matt. {\bf 33}, 475101 (2021).
%
\bibitem{Hos09} S. Hosokawa, M. Inui, Y. Kajihara, K. Matsuda, T. Ichitsubo, W.-C. Pilgrim, H. Sinn, L. E. González, D. J. González, S. Tsutsui, A. Q. R. Baron. Phys. Rev. Lett. {\bf 102}, 105502 (2009).
%
\bibitem{Hos13} S. Hosokawa, S. Munejiri, M. Inui, Y. Kajihara, W.-C. Pilgrim, Y. Ohmasa, S. Tsutsui, A. Q. R. Baron, F. Shimojo, K. Hoshino. J. Phys.: Condens. Matt. {\bf 25}, 
                112101 (2013).
%
\bibitem{Bry15} T. Bryk, G. Ruocco, T. Scopigno, A.P. Seitsonen. J. Chem. Phys. {\bf 143}
                104502 (2015).
%
\bibitem{Bry20} T. Bryk, T. Demchuk, J.-F. Wax, N. Jakse. J. Phys.: Condens. Matt. {\bf 32}, 
                184002 (2020).
%
\bibitem{Bry00} T. Bryk, I. Mryglod, J. Phys.: Condens. Matt. {\bf 12}, 6063 (2000).
%
\bibitem{Bry02} T. Bryk, I. Mryglod, J.Phys.:Condens.Matt. {\bf 14}, L445 (2002).
%
\bibitem{Kresse1996} Kresse G and Furthmüller J, Comput. Mater. Sci. 6 15–50 (1996).
%
\bibitem{SupI}  Supporting information
%
\bibitem{Behler2007} J. Behler and M. Parrinello, Phys. Rev. Lett. \textbf{98}, 146401 (2007).
%
\bibitem{Behler2021} J. Behler Chem. Rev. \textbf{121}, 10037 (2021).
%
\bibitem{Lammps} A. P. Thompson et al., Comput. Phys. Commun. 271, 108171 (2022); http://www.lammps.sandia.gov.
%
\bibitem{Singraber2019} A. Singraber, J. Behler, and C. Dellago, J. Chem. Theory Comput. \textbf{15}, 1827 (2019).
%
\bibitem{Jakse2023} N. Jakse, J. Sandberg, L. F. Granz, A. Saliou, P. Jarry, E. Devijver,  T. Voigtmann,  J. Horbach,  and A. Meyer, J. Phys.: Condens. Matter 51, 035402 (2023).
%
\bibitem{Demmel2025}  F. Demmel and N. Jakse, Phys. Rev. B \textbf{111}, L081104 (2025).
%
\bibitem{Sandberg2024} J. Sandberg, T. Voigtmann, E. Devijver, and N. Jakse, Mach. Learn.: Sci. Technol. 5, 025043 (2024).

\end{thebibliography}

\begin{thebibliography}{12}
	%
	\bibitem{Kre93} G.Kresse and J.Hafner, Phys. Rev.B {\bf 47}, 558 (1993);
	ibid. {\bf 49}, 14251 (1994).
	%
	\bibitem{Kre96} G.Kresse and J.Furthm\"uller, Comput. Mat. Sci. {\bf 6}, 15 (1996).
	%
	\bibitem{Kre96b} G.Kresse and J.Furthm\"uller, Phys. Rev.B {\bf 54}, 11169 (1996).
	\bibitem{Blo94} P.E. Bl\"ochl, Phys. Rev. B {\bf 50}, 17953 (1994).
	\bibitem{Kre99} G. Kresse, D. Joubert. Phys. Rev. B {\bf 59}, 1758 (1999).
	%
	\bibitem{Per96} J.P. Perdew, K. Burke, M. Ernzerhof.
	Phys. Rev. Lett. {\bf 77}, 3865 (1996).
	%
	\bibitem{Behler2011} J. Behler, J. Chem. Phys. \textbf{134}, 074106 (2011).
	%
	\bibitem{Jakse2023} N. Jakse, J. Sandberg, L. F. Granz, A. Saliou, P. Jarry, E. Devijver,  T. Voigtmann,  J. Horbach,  and A. Meyer, J. Phys.: Condens. Matter \textbf{51}, 035402 (2023).
	%
	\bibitem{Mry95} I.M. Mryglod, I.P. Omelyan, and M.V. Tokarchuk, Mol. Phys.
	{\bf 84}, 235 (1995).
	%
	%
	\bibitem{Bry10} T. Bryk, I. Mryglod, T. Scopigno, G. Ruocco, F. Gorelli, M. Santoro,                 J. Chem. Phys. {\bf 133}, 024502 (2010)
	%
	\bibitem{Bry13} T.Bryk, and G. Ruocco. Mol. Phys. {\bf 111}, 3457 (2013)
	
	\end{thebibliography}
\end{document}